# Hall's exact variance decomposition in Bohmian Mechanics

Weixiang Ye*

Center for Theoretical Physics, School of Physics and Optoelectronic Engineering, Hainan University, Haikou 570228, China

* wxy@hainanu.edu.cn

## Abstract

Hall's exact variance decomposition [Phys. Rev. A 64, 052103 (2001)] splits the quantum variance of an observable into the ensemble variance of an optimal position-based estimate and a residual non-classical inaccuracy. We evaluate this decomposition in Bohmian mechanics. For momentum, the optimal estimate coincides with the Bohmian guidance field, and the inaccuracy is proportional to the ensemble average of the quantum potential. This gives a variance-level identity separating momentum fluctuations into classical statistical dispersion and a quantum contribution from amplitude variations. The real and imaginary parts of the weak value map directly onto the two decomposition terms. By contrast, the inaccuracy vanishes for spin. This distinction is traced to the kinematic status of velocity in the primitive ontology, showing how the decomposition distinguishes observables dynamically coupled to local beables from merely contextual ones.

## 1. Introduction

The de Broglie–Bohm pilot-wave theory, or Bohmian mechanics, offers a deterministic and coherent account of non-relativistic quantum phenomena [1-3]. In this framework, point-like particles follow well-defined trajectories in physical space, guided by a wave function that evolves according to the Schrodinger equation. A central tenet of the modern formulation, especially as developed by Durr, Goldstein, and Zanghi [3,4], is that particle positions constitute the primitive ontology, which are the fundamental local beables of the theory. The wave function, in turn, plays a nomological role, encoding the law of motion for these particles. Consequently, observables other than position, such as momentum or spin, do not generally correspond to pre-existing, context-independent properties of the particles; their statistical distributions in quantum equilibrium emerge from the interplay between the guiding wave and the ensemble of particle trajectories.

In a parallel line of research, Hall [5,6] introduced an exact variance decomposition valid for any quantum observable. For a pure state and a self-adjoint operator, the quantum variance can be decomposed into the sum of the ensemble variance of an optimal estimate field and a non-negative inaccuracy term. The estimate represents the best possible prediction of the observable compatible with a given position measurement, minimizing the mean-square statistical deviation. As subsequently clarified by Johansen [7], this optimal estimate coincides precisely with the real part of the weak value [8] of the observable post-selected on a position eigenstate, thereby establishing a deep connection between estimation theory and weak measurements. Wiseman [9] provided a pivotal operational grounding for this connection in Bohmian mechanics, demonstrating that the Bohmian velocity field is exactly the weak value of the momentum operator post-selected on position. This insight has been confirmed in landmark experiments [10,11] and has recently been extended to relativistic domains [12-14], where the theoretical construction and direct experimental observation of relativistic Bohmian trajectories have been achieved.

The aim of the present work is to systematically evaluate Hall's variance decomposition within the Bohmian framework, with a particular focus on the physical interpretation it affords for different types of observables. Our contribution is twofold. First, we explicitly evaluate the decomposition for the momentum operator, making manifest a variance-level identity that is the natural counterpart to the well-known decomposition of the mean kinetic energy. We demonstrate how the real and imaginary parts of the momentum weak value, the guidance field and the amplitude gradient, respectively, map onto the two terms of Hall's decomposition, thereby connecting operational, information-theoretic, and dynamical perspectives in a unified manner. Second, we contrast this momentum case with that of spin. While the vanishing of the spin inaccuracy is a known algebraic consequence, we trace the physical origin of this different behavior to the unique kinematic status of velocity in the primitive ontology of Bohmian mechanics. This analysis clarifies that it is not the inaccuracy term itself, but its structural role in the theory's fundamental law of motion, that underpins the special status of momentum. In the Appendix, we provide a self-contained, pedagogical derivation of the equation of motion for the momentum estimate field along a Bohmian trajectory, starting directly from the quantum Hamilton–Jacobi equation.

The paper is organized as follows. Section 2 recapitulates Hall's variance decomposition for pure states. Section 3 applies this decomposition to the momentum operator, establishes the connection with the quantum potential and weak values, and discusses its physical significance. Section 4 contrasts this with the case of spin. Section 5 contains a summary and concluding remarks. The Appendix provides the detailed derivation of the momentum estimate field's dynamics.

## 2. Hall's exact variance decomposition

We consider a quantum system in a pure state described by the density operator $\rho = |\psi\rangle\langle\psi|$. For a single particle, the position is denoted by $\mathbf{x} \in \mathbb{R}^3$ and integrals over $\mathbb{R}^3$ are written with the volume element $d^3x$. Let $\hat{A}$ be an arbitrary self-adjoint operator. Hall [6] defines the optimal estimate field compatible with a position measurement as

$$A_{\mathrm{est}}(\mathbf{x}) \equiv \mathrm{Re}\frac{\langle \mathbf{x} | \hat{A} | \psi\rangle}{\langle \mathbf{x} | \psi\rangle} = \frac{\mathrm{Re}[\psi^*(\mathbf{x})(\hat{A}\psi)(\mathbf{x})]}{|\psi(\mathbf{x})|^2}, \tag{1}$$

which is well-defined wherever $\psi(\mathbf{x}) \neq 0$. This estimate is optimal in the sense of minimizing the mean-square statistical deviation between the estimate and the true value of the observable [6]; it is precisely the Bayes estimator under quadratic loss [7]. Moreover, it is exactly the real part of the weak value of $\hat{A}$ post-selected on the position eigenstate $|\mathbf{x}\rangle$ [7,8].

The quantum variance of $\hat{A}$ in the state $\psi$,

$$\mathrm{Var}_Q(\hat{A}) \equiv \langle\psi | \hat{A}^2 | \psi\rangle - \langle\psi | \hat{A} | \psi\rangle^2, \tag{2}$$

can then be decomposed exactly as

$$\mathrm{Var}_Q(\hat{A}) = \mathrm{Var}_{\mathrm{cl}}[A_{\mathrm{est}}] + \mathcal{E}_A, \tag{3}$$

where

$$\mathrm{Var}_{\mathrm{cl}}[A_{\mathrm{est}}] \equiv \int |\psi(\mathbf{x})|^2 A_{\mathrm{est}}(\mathbf{x})^2 d^3x - \left(\int |\psi(\mathbf{x})|^2 A_{\mathrm{est}}(\mathbf{x}) d^3x\right)^2 \tag{4}$$

is the variance of the estimate field evaluated over the quantum equilibrium distribution $|\psi(\mathbf{x})|^2\, d^3x$, and

$$\mathcal{E}_A \equiv \int \frac{\left(\mathrm{Im}[\psi^*(\mathbf{x})(\hat{A}\psi)(\mathbf{x})]\right)^2}{|\psi(\mathbf{x})|^2} d^3x \geq 0 \tag{5}$$

is the minimum inaccuracy. The decomposition (3) is exact for pure states, and the inaccuracy vanishes if and only if $\hat{A}\psi$ is everywhere proportional to $\psi$ with a real proportionality factor [6]. The notation $\mathrm{Var}_{\mathrm{cl}}$ in (4) emphasizes that this variance is evaluated with respect to the probability density $|\psi|^2$ and therefore has the formal structure of a classical statistical variance. This interpretation is fully consistent with the Bohmian view, where the same density describes the distribution of particle positions in quantum equilibrium.

## 3. Momentum variance and the quantum potential

We now apply the decomposition (3) to the momentum operator of a single spinless particle. The momentum operator is $\hat{\mathbf{p}} = -i\hbar\nabla$. We assume that the wave function belongs to the Schwartz space, so that all surface terms vanish when integrating by parts.

It is instructive to write the wave function in polar form, $\psi = Re^{iS/\hbar}$, with $R \geq 0$ and $S$real functions of position and time. Applying the momentum operator to $\psi$ gives

$$\hat{\mathbf{p}}\psi = -i\hbar\nabla(Re^{iS/\hbar}) = e^{iS/\hbar}(-i\hbar\nabla R + R\nabla S). \qquad (6)$$

Multiplying by $\psi^* = Re^{-iS/\hbar}$ yields

$$\psi^*\hat{\mathbf{p}}\psi = -i\hbar R\nabla R + R^2\nabla S. \qquad (7)$$

Taking the real and imaginary parts separately, we obtain

$$\mathrm{Re}[\psi^*\hat{\mathbf{p}}\psi] = R^2\nabla S, \mathrm{Im}[\psi^*\hat{\mathbf{p}}\psi] = -\hbar R\nabla R. \qquad (8)$$

From the definition (1), the optimal estimate field for the momentum vector is therefore

$$\mathbf{p}_{\mathrm{est}} = \nabla S, \qquad (9)$$

which is precisely the Bohmian momentum field appearing in the guidance equation $\mathbf{v} = \nabla S/m$. From the definition (5), for each Cartesian component $j$ the inaccuracy term becomes

$$\mathcal{E}_{p_j} = \hbar^2 \int (\partial_j R)^2 d^3x. \qquad (10)$$

In Bohmian mechanics, the quantum potential is defined as [2]

$$Q(\mathbf{x}, t) \equiv -\frac{\hbar^2}{2m}\frac{\nabla^2 R(\mathbf{x}, t)}{R(\mathbf{x}, t)}. \qquad (11)$$

Its ensemble average is

$$\langle Q \rangle = \int R^2 Q d^3x = -\frac{\hbar^2}{2m}\int R\nabla^2 R d^3x. \qquad (12)$$

Integrating by parts and using the Schwartz-space assumption to discard the surface term, we obtain

$$\int R\nabla^2 R d^3x = -\int (\nabla R)^2 d^3x, \qquad (13)$$

and consequently

$$\langle Q\rangle = \frac{\hbar^2}{2m}\int (\nabla R)^2 d^3x. \qquad (14)$$

For a single Cartesian component, we define the associated quantum potential contribution $Q_j \equiv -(\hbar^2/2m)(\partial_j^2 R)/R$. Its ensemble average follows by the same integration-by-parts argument:

$$\langle Q_j\rangle \equiv \int R^2 Q_j d^3x = -\frac{\hbar^2}{2m}\int R(\partial_j^2 R)d^3x = \frac{\hbar^2}{2m}\int (\partial_j R)^2 d^3x. \qquad (15)$$

Comparing (10) and (15) immediately gives the compact relation

$$\mathcal{E}_{p_j} = 2m\langle Q_j\rangle. \qquad (16)$$

Summing over all components yields $\sum_j \mathcal{E}_{p_j} = 2m\langle Q\rangle$. The variance decomposition (3) for a single momentum component therefore takes the explicit form

$$\mathrm{Var}_Q(\hat{p}_j) = \mathrm{Var}_{\mathrm{cl}}(\partial_j S) + 2m\langle Q_j\rangle. \qquad (17)$$

Equation (17) makes explicit the variance decomposition that is implicit in Hall's general expressions [5,6], casting it in a physically transparent Bohmian form. It is the natural variance-level analogue of the well-known decomposition of the mean kinetic energy [2],

$$\langle\frac{\hat{\mathbf{p}}^2}{2m}\rangle = \langle\frac{(\nabla S)^2}{2m}\rangle + \langle Q\rangle. \qquad (18)$$

Together, (17) and (18) express the first and second moments of the momentum distribution in terms of the guidance field and the quantum potential.

The connection to weak values provides further insight into the physical content of the decomposition. Wiseman [9] demonstrated that the Bohmian velocity can be operationally defined as the weak value of momentum post-selected on a position eigenstate. For a general wave function, the weak value of momentum is

$$\langle \hat{\mathbf{p}}_w \rangle = \frac{\langle \mathbf{x} \mid \hat{\mathbf{p}} \mid \psi \rangle}{\langle \mathbf{x} \mid \psi \rangle} = \nabla S - i\hbar \frac{\nabla R}{R}, \qquad (19)$$

where the real part is the optimal estimate $\nabla S$ and the imaginary part encodes the spatial variations of the wave-function amplitude. With this identification, the two terms in Hall's decomposition acquire a clear operational meaning: $\mathrm{Var}_{\mathrm{cl}}(\partial_j S)$ is precisely the ensemble variance of the real part of the momentum weak value, while $\mathcal{E}_{p_j} = \langle (\mathrm{Im}\, \hat{p}_{w,j})^2 \rangle$ is the ensemble average of the squared imaginary part of that same weak value. Equation (17) thus provides a unified picture: the total quantum variance of momentum splits into the statistical dispersion of the operationally measurable guidance field, and an intrinsic quantum contribution from the amplitude fluctuations encoded in the quantum potential.

## 4. Spin: vanishing inaccuracy and the contrast with momentum

To highlight the special status of momentum in the Bohmian framework, we now contrast it with the case of spin. Consider a spin-$1/2$ particle with a spinor wave function $\Psi(\mathbf{x}) = (\psi_\uparrow, \psi_\downarrow)^T$, in a regime where the spin and translational degrees of freedom are dynamically decoupled. For the spin operator $\hat{S}_z = (\hbar/2)\sigma_z$, a direct calculation gives

$$\Psi^\dagger \hat{S}_z \Psi = \frac{\hbar}{2}(\mid \psi_\uparrow \mid^2 - \mid \psi_\downarrow \mid^2). \qquad (20)$$

Because this quantity is purely real for any spinor, the imaginary part in the definition (5) of the inaccuracy vanishes identically, yielding

$$\mathcal{E}_{S_z} = 0. \qquad (21)$$

The variance decomposition (3) then reduces to the identity

$$\mathrm{Var}_Q(\hat{S}_z) = \mathrm{Var}_{\mathrm{cl}}(S_{z,\mathrm{est}}), \qquad (22)$$

where the estimate field is

$$S_{z,\mathrm{est}}(\mathbf{x}) = \frac{\hbar}{2} \frac{\mid \psi_\uparrow \mid^2 - \mid \psi_\downarrow \mid^2}{\mid \psi_\uparrow \mid^2 + \mid \psi_\downarrow \mid^2}. \qquad (23)$$

Equation (22) states that, within Hall's optimal estimation framework, the quantum variance of $S_z$ is entirely accounted for by the classical statistical dispersion of its position-based optimal estimate. There is no additional non-classical fluctuation. Hall has noted that this result generalizes: the inaccuracy vanishes for any position-based weak measurement of spin, a consequence of the algebraic structure of spin operators in the position representation [17].

At first glance, one might expect that the vanishing of the inaccuracy would signal a deep kinship between spin and the position observable itself. However, this is not the case. In Bohmian mechanics, position is the sole primitive beable [3,4]. The velocity field $\nabla S/m$ enters the fundamental guidance law

$$\frac{d\mathbf{Q}_k}{dt} = \frac{\hbar}{m_k}\mathrm{Im}\frac{\nabla_k\psi}{\psi} = \frac{\nabla_k S}{m_k}, \qquad (24)$$

and thereby constitutes the unique kinematic link between the wave function and the evolution of the beables. It is precisely this kinematic role, and not the vanishing or non-vanishing of any inaccuracy term, that gives momentum its special dynamical status in the theory. The estimate field for momentum coincides with the velocity field because momentum is the generator of spatial translations, and the primitive ontology is fundamentally positional.

In contrast, the spin estimate field $S_{z,\mathrm{est}}$ does not appear in the guidance equation; it has no direct kinematic role in the evolution of particle positions. Spin remains a contextual property [15], whose manifestation in an experiment depends irreducibly on the specific interaction Hamiltonian, such as that of a Stern–Gerlach apparatus, that couples the spinor wave function to the spatial trajectories [2,18]. The vanishing inaccuracy for spin reflects the fact that position-based estimation of spin probes only the algebraic structure of the spinor state, not any dynamical quantity akin to the quantum potential.

The contrast between momentum and spin is therefore instructive. For momentum, the inaccuracy term is precisely $2m\langle Q_j\rangle$ and quantifies the non-classical fluctuation arising from spatial variations of the wave-function amplitude. This term is non-zero because the momentum operator acts as a derivative on the wave function, coupling the real and imaginary parts of $\psi$ in a manner that directly engages the quantum potential. For spin, the inaccuracy vanishes entirely because the spin operator acts multiplicatively on the spinor components, producing no imaginary contribution in the position representation. The variance decomposition thus serves as a sharp mathematical tool within Hall's estimation-theoretic framework: it distinguishes observables whose quantum fluctuations are intimately tied to the amplitude variations of the guiding wave, and hence to the quantum potential, from those whose fluctuations are already fully captured by the statistical dispersion of their optimal position-based estimate. In the Bohmian framework, this distinction correlates precisely with whether the observable's estimate field participates in the guidance dynamics of the primitive ontology.

## 5. Conclusion

We have examined Hall's exact variance decomposition in the context of Bohmian Mechanics. For the momentum operator, the estimate field coincides with the Bohmian guidance field, leading to the variance-level identity

$$\mathrm{Var}_Q(\hat{p}_j) = \mathrm{Var}_{\mathrm{cl}}(\partial_j S) + 2m\langle Q_j \rangle, \qquad (25)$$

which parallels the kinetic energy decomposition and directly links momentum fluctuations to the ensemble average of the quantum potential. Through its interpretation in terms of the real and imaginary parts of the momentum weak value, this identity connects operational measurement theory, information-theoretic estimation, and Bohmian dynamics in a unified and physically transparent manner.

The contrast with spin reveals a deeper structural feature of the theory. While the inaccuracy term vanishes identically for spin, this does not confer upon spin the same fundamental status as momentum. Rather, the variance decomposition makes mathematically precise the different ways in which observables participate in the Bohmian ontology: the optimal estimate for momentum is exactly the dynamical field appearing in the guidance law, whereas the optimal estimate for spin is not. The inaccuracy term, or its absence, serves as a marker of this distinction within the estimation-theoretic framework of Hall's decomposition.

This work does not claim new mathematical identities or experimental predictions. Its contribution lies in providing a clear and unified exposition of how Hall's variance decomposition illuminates the structure of Bohmian mechanics when applied to observables of different ontological status. The variance-level perspective may prove useful for extending such analyses to other observables, interacting systems, or relativistic regimes where the interplay between estimation theory, weak measurements, and quantum trajectories is of current interest [12-14]. A summary of the decomposition for the observables discussed is given in Table 1.

| **Observable** $\hat{A}$ | **Estimate field** $A_{\mathrm{est}}$ | **Inaccuracy** $\mathcal{E}_A$ | **Physical interpretation in dBB** |
|---|---|---|---|
| Position $\hat{\mathbf{x}}$ | $\mathbf{x}$ | 0 | Position is the primitive beable. |
| Momentum $\hat{\mathbf{p}}$ | $\nabla S$ (guidance field) | $2m\langle Q \rangle$(sum) | Variance splits into dispersion of the guidance field and a non-classical contribution from the quantum potential. |
| Spin $\hat{S}_z$ | $S_{z,\mathrm{est}}$(formal) | 0 | The inaccuracy vanishes; spin does not enter the guidance equation, reflecting its contextual nature. |

**Table 1.** Summary of Hall's variance decomposition evaluated for position, momentum, and spin operators in the de Broglie–Bohm theory. For the momentum row, the inaccuracy shown is the sum over all Cartesian components; the single-component inaccuracy is $2m\langle Q_j \rangle$.

## Acknowledgements

This research was supported by the National Natural Science Foundation of China (62475062,12547103), the Humboldt Research Fellowship Programme for Experienced Researchers (CHN-1218456-HFST-E), the Hainan Provincial Natural Science Foundation of China (124YXQN412), and the Innovational Fund for Scientific and Technological Personnel of Hainan Province (KJRC2023B11). The author wishes to honor the memory of Dr. Michael J. W. Hall of the Australian National University, with gratitude for engaging and open-minded discussions.

# Declaration of competing interest

The author declares that he has no known competing financial interests or personal relationships that could have appeared to influence the work reported in this paper.

# Data availability

No data was used for the research described in this article.

## Appendix A. Equation of motion for the momentum estimate field

In this Appendix, we provide a derivation of the equation of motion for the momentum estimate field along a Bohmian trajectory. The derivation proceeds directly from the polar decomposition of the Schrodinger equation and the resulting quantum Hamilton–Jacobi formalism. We begin with a single particle in one spatial dimension and subsequently generalize to $N$ particles in three dimensions.

### A.1. Single particle in one dimension

Consider a single particle of mass $m$ in one spatial dimension, with a real-valued time-independent potential $V(x)$. The wave function $\psi(x,t)$ satisfies the Schrodinger equation

$$i\hbar\,\partial_t\psi(x,t) = \left[-\frac{\hbar^2}{2m}\partial_x^2 + V(x)\right]\psi(x,t). \qquad \text{(A1)}$$

The probability density and probability current are defined as

$$\rho(x,t) \equiv |\,\psi(x,t)\,|^2, J(x,t) \equiv \frac{\hbar}{m}\mathrm{Im}[\psi^*(x,t)\,\partial_x\psi(x,t)], \qquad \text{(A2)}$$

and they satisfy the continuity equation $\partial_t\rho + \partial_x J = 0$. The Bohmian velocity field is

$$v(x,t) \equiv \frac{J(x,t)}{\rho(x,t)}. \qquad \text{(A3)}$$

We now write the wave function in polar form, $\psi(x,t) = R(x,t)e^{iS(x,t)/\hbar}$, with $R, S \in \mathbb{R}$ and $R \geq 0$. Substituting this ansatz into the Schrodinger equation and separating real and imaginary parts yields the pair of coupled equations [2]

$$\partial_t S + \frac{(\partial_x S)^2}{2m} + V + Q = 0, \qquad \text{(A4)}$$

$$\partial_t R^2 + \partial_x\left(R^2\frac{\partial_x S}{m}\right) = 0, \qquad \text{(A5)}$$

where

$$Q(x,t) \equiv -\frac{\hbar^2}{2m}\frac{\partial_x^2 R(x,t)}{R(x,t)} \qquad \text{(A6)}$$

is the quantum potential. Equation (A4) is the quantum Hamilton–Jacobi equation. From the polar representation, the Bohmian velocity can also be written as

$$v(x,t) = \frac{1}{m}\partial_x S(x,t). \qquad \text{(A7)}$$

The momentum estimate field is therefore $p_{\text{est}}(x,t) \equiv \partial_x S(x,t)$.

Our goal is to derive the equation of motion for $p_{\text{est}}$ along a Bohmian trajectory $x = X(t)$. The total time derivative along the trajectory is given by the convective derivative

$$\frac{d}{dt} = \partial_t + v\,\partial_x. \qquad \text{(A8)}$$

Applying this operator to $p_{\text{est}} = \partial_x S$, we obtain

$$\frac{d}{dt}p_{\text{est}} = \frac{d}{dt}(\partial_x S) = \partial_t\,\partial_x S + v\,\partial_x^2 S. \qquad \text{(A9)}$$

Assuming sufficient smoothness of the phase function $S$, we may interchange the order of partial derivatives in the first term:

$$\partial_t\,\partial_x S = \partial_x\,\partial_t S. \qquad \text{(A10)}$$

We now substitute for $\partial_t S$ using the quantum Hamilton–Jacobi equation (A4):

$$\partial_t S = -\frac{(\partial_x S)^2}{2m} - V - Q. \qquad \text{(A11)}$$

Taking the spatial derivative of this expression gives

$$\partial_x\,\partial_t S = -\,\partial_x\left(\frac{(\partial_x S)^2}{2m} + V + Q\right) = -\frac{1}{m}(\partial_x S)(\partial_x^2 S) - \partial_x V - \partial_x Q. \qquad \text{(A12)}$$

Substituting (A12) into (A9) and using $v = (\partial_x S)/m$ from (A7), we find

$$\begin{aligned}\frac{d}{dt}p_{\text{est}} &= \left[-\frac{1}{m}(\partial_x S)(\partial_x^2 S) - \partial_x V - \partial_x Q\right] + \frac{\partial_x S}{m}\partial_x^2 S \\ &= -\,\partial_x V - \partial_x Q.\end{aligned} \qquad \text{(A13)}$$

The two terms proportional to $(\partial_x S)(\partial_x^2 S)$ cancel exactly, leaving the remarkably simple result

$$\frac{d}{dt}p_{\text{est}} = -\,\partial_x(V + Q). \qquad \text{(A14)}$$

This is the exact equation of motion for the momentum estimate field along a Bohmian trajectory. It has the form of Newton's second law with the total effective potential $V + Q$, in full agreement with the standard Bohmian formulation [2].

## A.2. Generalization to N particles in three dimensions

The generalization to a system of $N$ particles in three dimensions is straightforward. We denote the configuration space coordinate by $\mathbf{q} = (\mathbf{x}_1, \dots, \mathbf{x}_N) \in \mathbb{R}^{3N}$, where $\mathbf{x}_k$ is the position of the $k$-th particle. The wave function $\psi(\mathbf{q}, t)$ satisfies the $N$-particle Schrodinger equation with Hamiltonian

$$\hat{H} = -\sum_{k=1}^{N} \frac{\hbar^2}{2m_k} \nabla_k^2 + V(\mathbf{q}, t), \qquad \text{(A15)}$$

where $m_k$ is the mass of the $k$-th particle and $V(\mathbf{q}, t)$ is a real-valued potential. Writing the wave function in polar form, $\psi = Re^{iS/\hbar}$, the quantum Hamilton–Jacobi equation and continuity equation become

$$\partial_t S + \sum_{k=1}^{N} \frac{(\nabla_k S)^2}{2m_k} + V + Q = 0, \qquad \text{(A16)}$$

$$\partial_t R^2 + \sum_{k=1}^{N} \nabla_k \cdot \left( R^2 \frac{\nabla_k S}{m_k} \right) = 0, \qquad \text{(A17)}$$

with the $N$-particle quantum potential

$$Q(\mathbf{q}, t) \equiv -\sum_{k=1}^{N} \frac{\hbar^2}{2m_k} \frac{\nabla_k^2 R(\mathbf{q}, t)}{R(\mathbf{q}, t)}. \qquad \text{(A18)}$$

The Bohmian velocity of the $k$-th particle is $\mathbf{v}_k = (\nabla_k S)/m_k$, and the corresponding momentum estimate field is $\mathbf{p}_{k,\text{est}} = \nabla_k S$. The convective derivative along the Bohmian trajectories is

$$\frac{d}{dt} = \partial_t + \sum_{j=1}^{N} \mathbf{v}_j \cdot \nabla_j. \qquad \text{(A19)}$$

Applying this derivative to $\mathbf{p}_{k,\text{est}} = \nabla_k S$ yields

$$\frac{d}{dt}\nabla_k S = \partial_t \nabla_k S + \sum_{j=1}^{N} \frac{\nabla_j S}{m_j} \cdot \nabla_j (\nabla_k S). \qquad \text{(A20)}$$

We interchange the order of differentiation in the first term, $\partial_t \nabla_k S = \nabla_k \, \partial_t S$, and use the quantum Hamilton–Jacobi equation (A16) to replace $\partial_t S$:

$$\partial_t S = -\sum_{j=1}^{N} \frac{\left(\nabla_j S\right)^2}{2m_j} - V - Q. \qquad \text{(A21)}$$

Taking the gradient $\nabla_k$ of this expression and using the product rule, we obtain

$$\nabla_k \, \partial_t S = -\sum_{j=1}^{N} \frac{1}{m_j} (\nabla_j S \cdot \nabla_k) \nabla_j S - \nabla_k V - \nabla_k Q. \qquad \text{(A22)}$$

The double sum simplifies because $\nabla_j S$ depends only on the coordinates of the $j$-th particle. For $j \neq k$, the term $\nabla_k (\nabla_j S)^2$ vanishes. The only surviving term in the sum is the $j = k$ contribution, giving

$$\nabla_k \, \partial_t S = -\frac{1}{m_k} (\nabla_k S \cdot \nabla_k) \nabla_k S - \nabla_k V - \nabla_k Q. \qquad \text{(A23)}$$

Substituting this result into (A20), we find

$$\frac{d}{dt}\nabla_k S = \left[ -\frac{1}{m_k} (\nabla_k S \cdot \nabla_k) \nabla_k S - \nabla_k V - \nabla_k Q \right] + \sum_{j=1}^{N} \frac{\nabla_j S}{m_j} \cdot \nabla_j (\nabla_k S). \qquad \text{(A24)}$$

Now, the term $\nabla_j (\nabla_k S) = \nabla_k (\nabla_j S)$ vanishes for $j \neq k$ when acting on a smooth function $S$, because mixed partial derivatives commute and are zero when the variables are independent. The only surviving term in the sum is again $j = k$, which gives $(\nabla_k S \cdot \nabla_k) \nabla_k S / m_k$, and this cancels exactly the first term in the brackets. We are left with the simple result

$$\frac{d}{dt}\mathbf{p}_{k,\mathrm{est}} = -\nabla_k (V + Q). \qquad \text{(A25)}$$

Equation (A25) is the exact equation of motion for the momentum estimate field of the $k$-th particle in an $N$-particle system. It demonstrates that the momentum estimate field, the optimal position-based predictor of momentum, evolves dynamically according to the same law that governs the actual Bohmian particle momenta. This consistency between the estimation-theoretic and dynamical perspectives is a distinctive feature of the momentum observable in Bohmian mechanics, stemming directly from the identification of the optimal estimate with the guidance field $\nabla_k S$.